# Designing superhard magnetic material in clathrate *β*-C$_3$N$_2$ through atom embeddedness[*]

Liping Sun[a], Botao Fu[b] and Jing Chang[*][a]

[a] College of Physics and Electronic Engineering, Sichuan Normal University, Chengdu, 610068, China

[b] College of Physics and Electronic Engineering, Center for Computational Sciences,

Sichuan Normal University, Chengdu, 610068, China

**Abstract:** Designing new compounds with the coexistence of diverse physical properties is of great significance for broad applications in multifunctional electronic devices. In this work, based on density functional theory, we predict the coexistence of mechanical superhardness and the controllable magnetism in the clathrate material *β*-C$_3$N$_2$ through the implant of the external atom into the intrinsic cage structure. Taking hydrogen-doping (H@*β*-C$_3$N$_2$) and fluorine-doping (F@*β*-C$_3$N$_2$) as examples, our calculations indicate these two doped configurations are stable and discovered that they belong to antiferromagnetic semiconductor and ferromagnetic semi-metal, respectively. These intriguing magnetic phase transitions originate from their distinctive band structure around the Fermi level and can be well understood by the 3D Hubbard model with half-filling occupation and the Stoner model. Moreover, the high Vickers hardness of 49.0 GPa for H@*β*-C$_3$N$_2$ and 48.2 GPa for F@*β*-C$_3$N$_2$ are obtained, suggesting they are clathrate superhard materials as its host. Therefore, the incorporation of H and F in *β*-C$_3$N$_2$ gives rise to a new type of superhard antiferromagnetic semiconductor and superhard ferromagnetic semimetal, respectively, which could have potential applications in harsh conditions. Our work provides an effective strategy to design a new class of highly desirable multifunctional materials with excellent mechanical properties and magnetic properties, which may arouse spintronic applications in superhard materials in the future.

## 1. Introduction

A clathrate is a large group of compounds consisting of cage-like lattice structures that can trap guest atoms or molecules. Due to the variable combination of possible guest atoms and host polyhedral cages, clathrates exhibit tunable physical properties and have received significant attention in condensed matter physics and materials science.[0] For instance, a DFT-predicted novel clathrate boride, made up of face-sharing B$_{26}$ cages encapsulating a single La atom, demonstrates phonon-mediated superconductivity of 18 K at ambient pressure,[2] where the transition temperature can be significantly improved by substituting various guest atoms inside the cage.[3-5]

---

[*] *E-mail:* changjing0394@163.com.

Moreover, the incorporation of cerium as a guest atom into the intermetallic clathrate is reported, leading to exceedingly low lattice thermal conductivities and remarkable enhancement of thermopower with respect to a rare-earth-free reference material.[6,7] Besides, the robust ferroelectricity with high transition temperature is recently observed in a new clathrate with a carbon-boron framework, opening the possibility for a new class of ferroelectric materials with potential across a broad range of applications.[8]

On the other hand, superhard materials, identified by a high Vickers hardness ($H_V$) of more than 40 GPa,[9] are indispensable to fundamental study in condensed-matter physics and have been practically applied to a wide range of technical fields, such as cutting and polishing tools, wear-resistant coatings, etc. In recent years, several clathrate materials such as carbon allotrope,[10,11] sodalite-like born-carbon,[12] and boron-nitrogen framework,[13,14] are reported as potential superhard materials. These clathrate superhard materials are made up of relatively small cages formed by strong covalent bonds from light elements (B, C, and N). Distinct from traditional superhard materials with uniformly insulating nature, new clathrate superhard materials are endowed with tunable and multiple functions due to their inner cages that can accommodate various guest atoms. For example, the synthesis of the superhard iron tetraboride superconductor opens a new class of highly desirable clathrate materials combining advanced mechanical properties and superconductivity,[15,16] Therefore, searching and designing for new clathrate superhard materials combined with other comprehensive properties such as ferromagnetism, ferroelectricity, thermoelectricity, and superconductivity[17-19] are of great significance for various applications in science and engineering.

In particular, the combination of magnetism with superior mechanical hardness, to fabricate superhard magnetic materials, has garnered significant research interest[20-22], which could have potential applications in harsh conditions, such as micro-electromechanical systems, magnetic actuators, and magnetic sensors in high-speed motors. However, only a few compounds such as transition metal borides[23], have been reported to possess both magnetism and high hardness. Here in this paper, based on the flexible adjustability of clathrate materials, we propose an alternative strategy to realize the coexistence of magnetism and superhardness in doped clathrate compounds. Starting from the previously predicted clathrate superhard materials $\beta$-$C_3N_2$,[24,25] our calculations indicate that the intersection of hydrogen and fluorine atoms might induce antiferromagnetic semiconductor and ferromagnetic metal states, and meanwhile maintain the high Vickers hardness of 49 GPa and 48.2 GPa, respectively. The underly mechanisms of super-high hardness and different magnetic ground states are revealed through the analysis of bonding characteristics, charge transfer, and mechanical constants. Thereinto, the H-doping system (H@$\beta$-$C_3N_2$) belongs to 3D Hubbard model with half-filling, which might spontaneously transform into an antiferromagnetic semiconductor ground state with a large U limit, while in F doped system (F@$\beta$-$C_3N_2$) the strong orbital hybridization between guest and host atoms gives rise to a large DOS peak at Fermi level which further generates a Stoner ferromagnetic metal due to electronic instability. Therefore, the incorporation of H and F in $\beta$-$C_3N_2$ might give rise to a new type of superhard antiferromagnetic semiconductor and superhard ferromagnetic semimetal, respectively. Our work provides an effective strategy to design a new class of highly desirable multifunctional materials with excellent mechanical properties and magnetic properties.

## 2. Computational methods

The first-principles calculations were performed using the Vienna ab initio simulation package (VASP).[26] The Perdew-Burke-Ernzerhof (PBE) functional in the generalized gradient approximation was used.[27] The projector augmented wave (PAW) pseudopotential considers $2s^22p^2$ and $2s^22p^3$ as valence electrons for C and N atoms, respectively. A plane-wave energy cutoff was set at 800 eV and the Brillouin−zone (BZ) integration is carried out using a 7×7×7 Monkhorst−Pack grid in the first BZ for relaxation. During the geometrical optimization, all forces on atoms converged to less than 0.01eV/Å and the change of the total energy is less than $10^{-5}$ eV. Before calculating the band structure, the self-consistent calculation was carried out with a much denser k-point grid by 10×10×10 to obtain a more accurate charge density and band structure ultimately. Phonon calculation was performed with 2×2×2 supercell by using PHONOPY code.[28] Elastic constants were calculated by the strain-stress method [29] and the bulk modulus, shear modulus, and Poisson's ratio of these structures were derived from the Voigt–Reuss–Hill approximation.[30]

## 3. Geometric structure and stability

Recently, a group of clathrate carbon-nitride compounds with superhard characteristics were proposed. Thereinto, $\beta$-$C_3N_2$ with space group of P-43m belongs to a sodalite-like cage structure that is composed of inter-linked strong C-N bonds, which provides a new degree for the manipulations of physical properties. As shown in Fig.1 (a), $\beta$-$C_3N_2$ crystals in a cubic lattice with an ultimately optimized lattice constant of 5.084 Å, where twelve carbon atoms located on Wyckoff positions of 12$i$ (0.156, 0.156, 0.497) and eight nitrogen atoms at 4$e$ (0.246, 0.246, 0.246) and 4$e$ (0.731, 0.731, 0.731). The average radius of the cage is about 1.30 Å, which allows for the intercalation of light elements with a proper atomic radius. Therefore, we consider these elements in the first three periods and the group VIIA in the periodic tables and finally obtained several stable structures with various electronic structure and magnetic properties [see supplementary material]. Here we take the hydrogen ($r$=0.53 Å) and fluorine ($r$=0.71 Å) as examples to explore potentially rich physical properties induced by external intercalation.

We first considered possible doping sites for inserting of H and F into cage-like β-$C_3N_2$ and found that the central position (0.5, 0.5, 0.5) is locally energy stable for H[31] and F doping, and robust against perturbations. The corresponding discussions are given in detail in the supplementary materials. In the main body, our primary focus is on the H- and F- center doped crystal structures, as shown in Fig. 1, which are named H@$β$-$C_3N_2$ and F@$β$-$C_3N_2$, respectively. The lattice constants of H@$β$-$C_3N_2$ and F@$β$-$C_3N_2$ are slightly increased by 0.017 Å and 0.038 Å compared with pristine $β$-$C_3N_2$. The phonon spectra of pristine $β$-$C_3N_2$, and H and F doped $β$-$C_3N_2$ at 0 GPa were investigated with 2×2×2 supercell, these results are shown in Figs. 1(d)-(f). It is obvious that there is no imaginary frequency in the whole Brillouin zone for H@$β$-$C_3N_2$ and F@$β$-$C_3N_2$, which indicates that they are dynamically stable at 0 GPa. In addition, the phonon dispersions for two doped systems are almost the same as that of their parent $β$-$C_3N_2$, implying excellent mechanical properties still be remained in the doped systems.

Furthermore, thermodynamic stability has been confirmed by the calculation of cohesive energy through the chemical equation: $ΔH_c = E_{total}$ (X@$C_3N_2$)-$3E_C$-$2E_N$-$E_X$, where $E_{total}$ (X@$C_3N_2$) is the total energy of compound, $E_C$ is the energy of C atom, $E_N$ is the energy of N atom, $E_X$ is the energy of the doped atom. The calculated negative cohesive energy of -4.202, -4.002 and -3.941 eV/unit cell for $β$-$C_3N_2$, H@$β$-$C_3N_2$ and F@$β$-$C_3N_2$, respectively, shows the energy release during compound formation, which also confirms the thermodynamic stability of the studied materials. To further verify the thermodynamic stability of H@$β$-$C_3N_2$ and F@$β$-$C_3N_2$, we have performed Ab initio molecular dynamics simulations (AIMD) at room temperature. A 2 × 2 × 2 supercell with a time step of 1 fs during the simulation was used. The calculation results are shown in Figs. 1 (g)-(i), which indicates that both H@$β$-$C_3N_2$ and F@$β$-$C_3N_2$ remain dynamic stability at 300 K.

## 4. Electronic structures

To further insight into the properties of these doped compounds, the electronic structures of pristine $β$-$C_3N_2$, H and F doped $β$-$C_3N_2$ are calculated and shown in Figs. 2 (a)-(c), respectively. It is found that the pristine $β$-$C_3N_2$ is a wide indirect band gap ($E_g$=3.564 eV) insulator, with the valence-band maximum located at Γ point and the conduction-band minimum located at M point. We find that the highest valence bands are almost composed of N-$p$ orbitals while the lowest conduction bands mainly derive from C-$p$ orbitals with a small number of N-$p$ orbitals. For H@$β$-$C_3N_2$ in Fig. 2 (b), a nearly isolated band appears inside the wide band gap of pristine $β$-$C_3N_2$, which crosses the Fermi level, and gives an extended Fermi surface crossing the Brillouin zone boundary as shown in Fig. 2(e). To reveal the origin of the unique band around the Fermi level of H@$β$-$C_3N_2$, we calculated band projection, partial density of state (PDOS) and charge density. From the band projection and PDOS, we find this unique band is mainly composed of H's s-orbital, which is consistent with the charge density plotted in Fig. 2 (h), where electrons are mainly located on H-atoms in real space. Further, via the Bader charge analysis in Fig. S7, we found the inserted H-atom only gains 0.028 electrons, which indicates the interaction between the doped H-atom and $\beta$-C$_3$N$_2$ is very weak. Since there is only one H atom in the simple cubic cell, this isolated *s*-band has to be half-filled by one electron from H atom. Therefore, it can be described as a single-bond Hamiltonian in the cubic lattice made up of H-atom,

$$H = E_0 - t \sum_{\langle ij \rangle} C_{i\sigma}^+ C_{j\sigma} + H.C. \tag{1}$$

The $E_0$ is onsite energy and $t$ is the nearest neighbor hopping. The energy spectrum is analytically given as, $E(\mathbf{k})=E_0-2t(\cos k_x+\cos k_y+\cos k_z)$, which gives a similar constant energy surface as shown in Fig. 2 (g). It's worthwhile considering the electron-electron interaction (Hubbard U) belongs to a standard three-dimensional Hubbard model in a simple cubic lattice, which contains rich phase diagrams as pointed out by previous studies.[32]

Compared with the H@$\beta$-C$_3$N$_2$, the band structure of the F@$\beta$-C$_3$N$_2$ system is more complicated. As exhibited in Fig. 2(c), there is a conductive band that crosses the Fermi level, giving rise to the Fermi pockets centered at Γ and M points as displayed in Fig. 2 (f). However, this band is entangled with other two valence bands, which becomes triple degenerate at the R point. From the band projection and PDOS, we find these entangle bands are almost equally composed of F's p-orbital and N's p-orbital, indicating strong orbital hybridization between inserted-F and pristine $\beta$-C$_3$N$_2$, which is consistent with the charge density plotted in Fig. 2 (i) where electrons are mainly located on both F-atom and N-atom in real space. In addition, via the Bader charge analysis in Fig. S7, we find the inserted F-atom gains 0.573 electrons, which indicates the interaction between the doped F-atom and $\beta$-C$_3$N$_2$ is rather large. Intriguingly, the relatively dispersionless bands with saddle points lead to a peak of DOS at the Fermi level, which may further contribute to remarkably correlated effects and novel physical properties.[33] As a whole, both H and F-doping will induce a metallic state in $\beta$-C$_3$N$_2$.

## 5. Magnetic properties

The rich physical phenomena could be induced by doping atoms in cage-like materials, such as ferroelectricity and magnetism. With the unusual half-filled and flat-band characteristics of H and F doped $\beta$-C$_3$N$_2$, we further explore the underlying magnetic phase transition of these systems. The spin-polarized calculation is performed which partially includes the exchange-correlation effect, and then we find both H@$\beta$-C$_3$N$_2$ and F@$\beta$-C$_3$N$_2$ have non-zero net magnetic moments, which indicates magnetic phases are energy favorable. The magnetic moment distribution in real space is directly reflected by the illustration of effective spin charge density ($\Delta\rho =\rho\uparrow-\rho\downarrow$) in Fig. 3, where a magnetic moment of 1.00 $\mu_B$ is mainly localized on H-atom for H@$\beta$-C$_3$N$_2$, while a magnetic moment of 0.91 $\mu_B$ magnetic moment is extensively distributed on both F-atom and N-atom. These phenomena are consistent with the analysis of the partial density of states and charge density distribution around the Fermi level. In order to probe the most stable magnetic configurations, a $\sqrt{2} \times \sqrt{2} \times 2$ supercell is built with four types of magnetic configurations considered in Figs. 4(a)-(d). It involves a ferromagnetic (FM) phase and three antiferromagnetic (AFM) phases such as type-A, -C, -G AFM. The final total energies for these configurations are listed in Table 1, with respect to the nonmagnetic (NM) phases. One can find that the H@$\beta$-C$_3$N$_2$ has the type-G AFM ground state while the F@$\beta$-C$_3$N$_2$ possesses the FM ground state.

The spin-polarized electronic structures of type-G AFM H@$\beta$-C$_3$N$_2$ and of FM F@$\beta$-C$_3$N$_2$ are presented in Fig. 4(e) and 4(f), respectively. Thereinto, the H@$\beta$-C$_3$N$_2$ exhibits AFM semiconducting nature with a band gap of 2.485 eV. This magnetic phase transition from normal half-filled metal into AFM semiconductor can be well understood by the 3D Hubbard model for a simple cubic lattice,[34,35] when a large coupling limit (U/t) is satisfied in H@$\beta$-C$_3$N$_2$. On the contrary, the F@$\beta$-C$_3$N$_2$ exhibits an FM metal state as shown in Fig. 4(f), where the spin-down bands move up with the triplet point coincidently locate at the Fermi level, while the spin-up bands move down with the band around Γ point crossing the Fermi level. From the spin-splitting DOS, we noticed that the Fermi level is mainly occupied by the spin-down channel, giving rise to a high spin polarization of 80%. With only 0.03 electron per unit cell doping per unit cell ($0.263 \times 10^{-3}$ e/Å$^3$), the F@$\beta$-C$_3$N$_2$ would transform into 100% spin polarization, namely FM half-metal. Different from the AFM ground state of H@$\beta$-C$_3$N$_2$, the F@$\beta$-C$_3$N$_2$ experiences a phase transition from a normal metal state into FM metal. This transition would be comprehended by the Stoner criterion for itinerant ferromagnetism:[36] $N(E_F)I > 1$, where $N(E_F)$ is the DOS at the Fermi energy in the NM state and $I$ is the stoner parameter. Here the almost divergent DOS peak at the Fermi level in Fig. 2(c) makes the Stoner criterion to be satisfied, [37] and the ferromagnetic state is naturally induced with lower total energy. Moreover, from the PDOS of F@$\beta$-C$_3$N$_2$, we further identify that the itinerant ferromagnetism originates from the strong $p$-orbital hybridization of F and N atoms due to larger and extended $2p$ orbital. This is also revealed by the distribution of magnetic moment that the N-atoms have 0.466 $\mu_B$ while the F-atom possesses 0.293 $\mu_B$. According to the above analysis, H@$\beta$-C$_3$N$_2$ and F@$\beta$-C$_3$N$_2$ are confirmed to be antiferromagnetic semiconductor and ferromagnetic metal, respectively. The magnetism in doped $\beta$-C$_3$N$_2$ mainly derived from the $p$-orital of inserted atom, which is different from present magnetic superhard materials [22, 23] (such transition metal boride) in which the magnetism mainly originates from the unpaired $d$-orbital of transition atoms.

## 6. Elastic constants and mechanical properties

The atom intercalation not only dramatically modify the electronic structure of hosts but can also remarkably affect the mechanical properties. On account of the expectation of designing superhard crystal the mechanical property of doping $\beta$-$C_3N_2$ is investigated inevitably. The elastic behavior is of key importance for us to understand the deformation behavior for superhard materials in response to external forces. The calculated elastic constants are listed in Table 2. For a cubic structure, there are three independent elastic constants including $C_{11}$, $C_{12}$, and $C_{44}$. To be a stable cubic crystal, the elastic constant must obey the following mechanical criteria: [38] $C_{11} - C_{12} > 0$, $C_{11} + C_{12} > 0$, $C_{44} > 0$. From these calculated values of elastic constants in Table 2, it is obvious that these elastic constants are positive and satisfy the mechanical criteria, revealing that both H@$\beta$-$C_3N_2$ and F@$\beta$-$C_3N_2$, are mechanically stable.

The Vickers hardness of materials, $H_v$ (GPa), can be obtained according to the empirical model: [39]

$$H_V = 2(k^2 G)^{0.585} - 3 \qquad (2)$$

where $k = G/B$. Thereinto, the bulk modulus ($B$) and shear modulus ($G$) show the resistance to fracture and plastic deformation that can be directly derived from elastic constants by the Voigt-Reuss-Hill approximation. [30] The Pugh's ratio defined as $B/G$, similar to Poisson's ratio $\nu$, can be used to characterize the brittleness or ductility of a material. Generally speaking, solid with $B/G < 1.75$ and $\nu < 0.26$ refers to material behaving in a brittle manner, otherwise, it corresponds to ductility.[40] In general, brittle materials with smaller $B/G$ tend to have higher hardness. From Table 2, the calculated $B/G$ and Poisson's ratio are 1.11 and 0.150 for H@$\beta$-$C_3N_2$, 1.11 and 0.156 for F@$\beta$-$C_3N_2$, which are comparable with that of pristine $\beta$-$C_3N_2$ (0.93 and 0.105). All these results indicate that these two doped systems still maintain favorable brittleness as their parent $\beta$-$C_3N_2$. As a result, the Vickers hardness based on formula (2) is further obtained. The obtained values of $H_v$ are 49.0 GPa and 48.2 GPa for H@$\beta$-$C_3N_2$ and F@$\beta$-$C_3N_2$, respectively. Although these values are a little smaller than that of pristine $\beta$-$C_3N_2$ (58.2 Gpa), they are higher than 40 GPa, which suggests that both H@$\beta$-$C_3N_2$ and F@$\beta$-$C_3N_2$ are superhard materials as expected. Besides, in order to reveal the origin of their excellent mechanical properties, we also calculated the electron localization functions as shown in Fig. S7 in supplementary materials, from which one can identify that it is the strong B-N covalent bonding is responsible for the superhigh hardness in both pristine and doped $\beta$-$C_3N_2$. Hence, based on the above results, we confirmed the coexistence of magnetism and superharness in doped $\beta$-$C_3N_2$, providing potential material candidates for superhard magnetic materials.

## 7. Conclusions

We have shown that the incorporation of H and F in $\beta$-$C_3N_2$ can realize the combination of excellent mechanical properties and distinctive magnetic properties. Via first-principles calculation, we clearly reveal it is the unique electronic structure from the intercalation atoms around the Fermi level that lead to the AFM semiconductor state and FM semimetal state in H@$\beta$-$C_3N_2$ and F@$\beta$-$C_3N_2$, respectively. Meanwhile, the high Vicker hardness, as well as brittleness, still remained in both doped systems. Therefore, we think the intercalation of proper atoms in existing clathrate superhard materials can effectively induce magnetism and manipulate the electronic structure. This work extends the magnetic property of superhard materials and maybe can arouse spintronic applications in superhard materials.

## Acknowledgements

This work was supported by the National Natural Science Foundation of China (Grant No. 12204330) and the Sichuan Normal University for financial support (No. 341829001).

Table 1 Calculated total energy of NM, FM and three AFM states (type-A, type-C, type-G). The energy of non-magnetic state was set as a reference standard. It should be noted that the AFM states of F@$\beta$-C$_3$N$_2$ are unstable which will spontaneously converge to the NM state.

| Energy (meV) | NM | FM | FM | Type-C AFM | Type-G AFM |
| --- | --- | --- | --- | --- | --- |
| H@$\beta$-C$_3$N$_2$ | 0 | -342.98 | -360.40 | -375.46 | -391.97 |
| F@$\beta$-C$_3$N$_2$ | 0 | -16.50 | — | — | — |

Table 2 Calculated elastic constants $C_{ij}$ (GPa), bulk modulus $B$ (GPa), shear modulus $G$ (GPa), $B/G$ ratio, Poisson's ratio ($\nu$) and $H_v$ (GPa) of $\beta$-C$_3$N$_2$, H@$\beta$-C$_3$N$_2$ and F@$\beta$-C$_3$N$_2$.

| | $C_{11}$ | $C_{12}$ | $C_{44}$ | $B$ | $G$ | $B/G$ | $\nu$ | $H_v$ |
| --- | --- | --- | --- | --- | --- | --- | --- | --- |
| $\beta$-C$_3$N$_2$ | 834 | 98 | 329 | 343 | 368 | 0.93 | 0.105 | 58.2 |
| H@$\beta$-C$_3$N$_2$ | 841 | 131 | 314 | 368 | 330 | 1.11 | 0.150 | 49.0 |
| F@$\beta$-C$_3$N$_2$ | 805 | 149 | 328 | 368 | 328 | 1.12 | 0.156 | 48.2 |

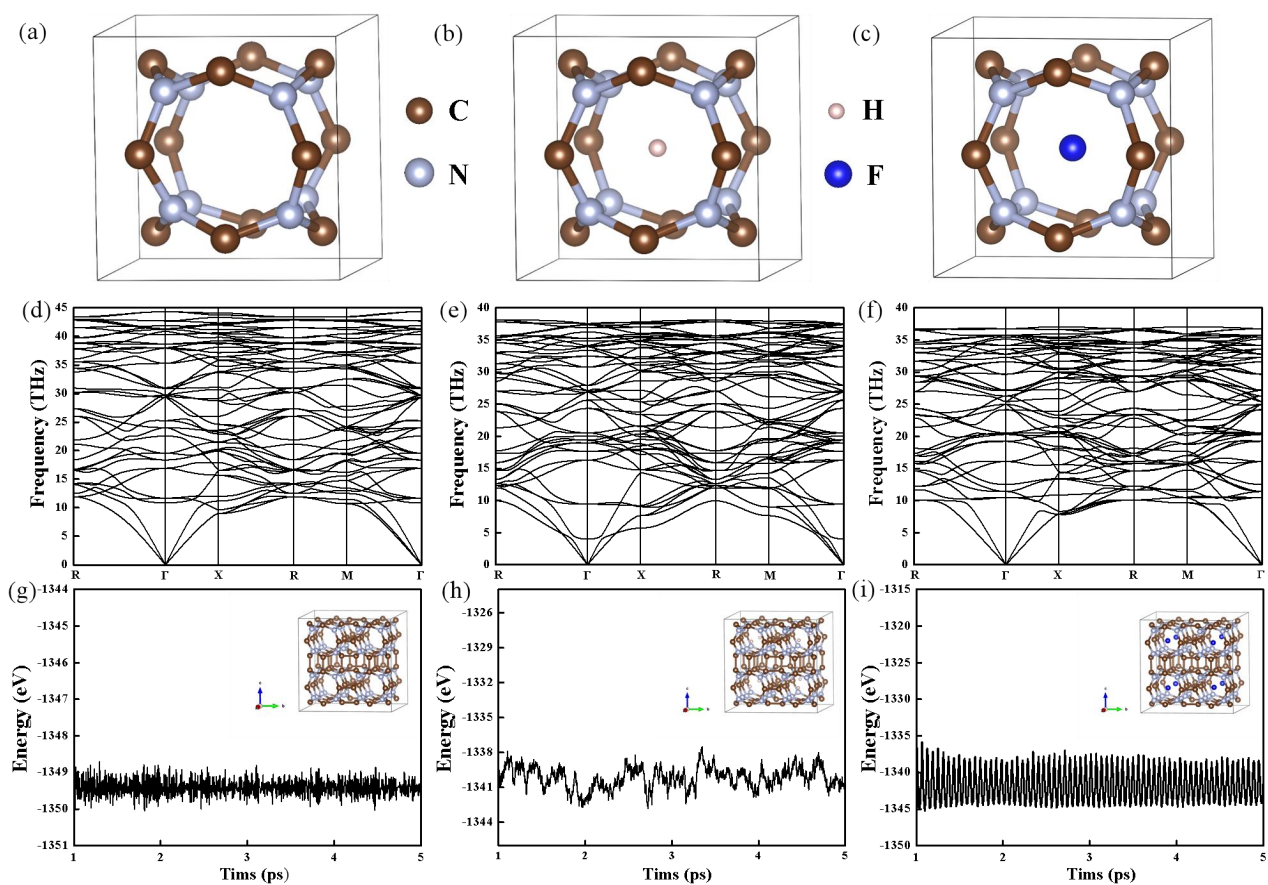

Fig. 1 Crystal structure of (a) pristine $\beta$-$C_3N_2$, (b) H@$\beta$-$C_3N_2$, and (c) F@$\beta$-$C_3N_2$. Phonon dispersion of (d) $\beta$-$C_3N_2$, (e) H@$\beta$-$C_3N_2$, and (f) F@$\beta$-$C_3N_2$. AIMD simulation under 300 K of the (g) $\beta$-$C_3N_2$, (h) H@$\beta$-$C_3N_2$, and (i) F@$\beta$-$C_3N_2$, the crystals after 5 ps are given.

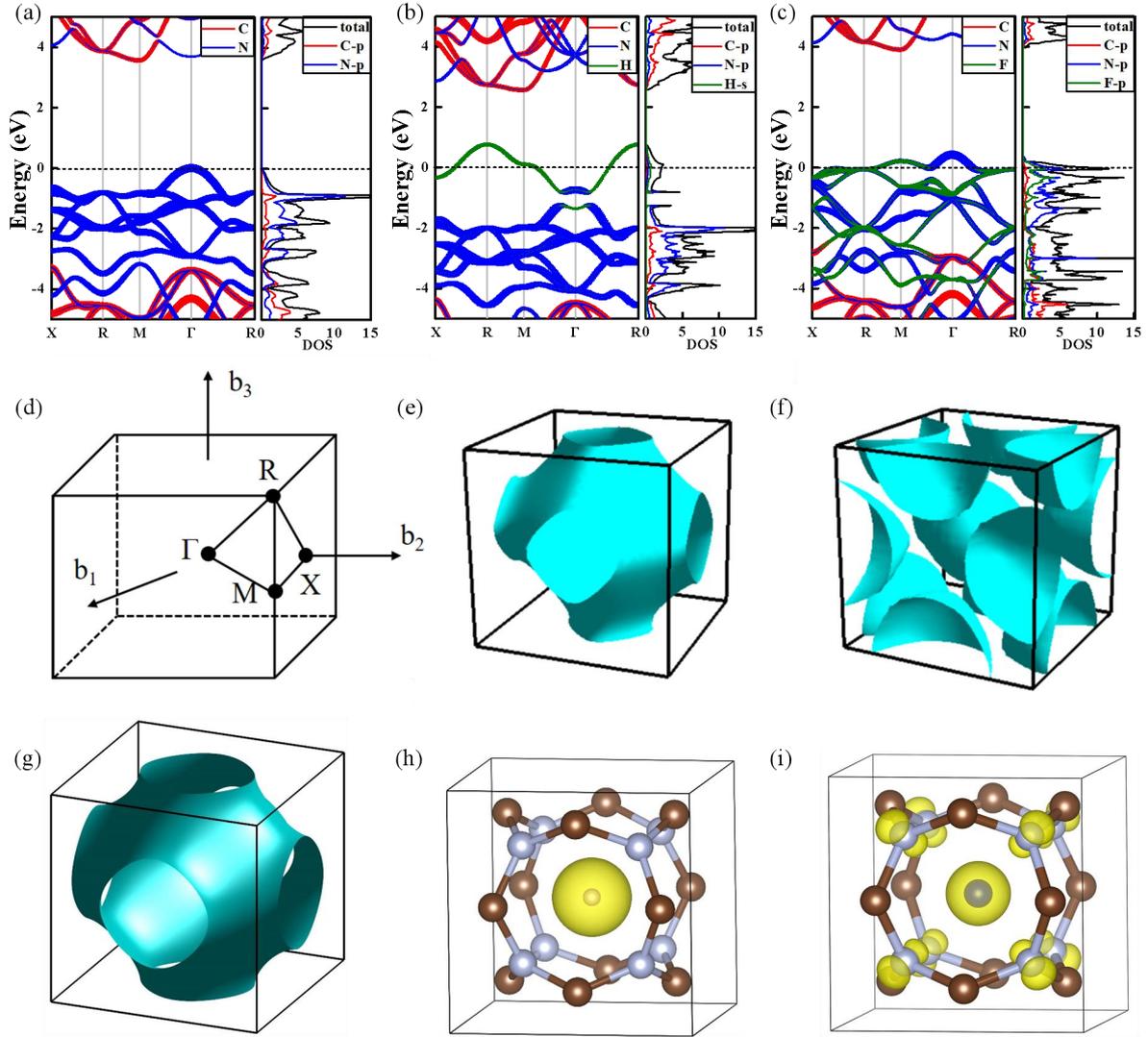

Fig. 2 Band structure and density of states of (a) $\beta$-C$_3$N$_2$, (b) H@$\beta$-C$_3$N$_2$ and (c) F@$\beta$-C$_3$N$_2$, respectively. (d) The first Brillouin zone for a simple cubic crystal with high-symmetric points. (e) and (f) are Fermi surfaces from for first-principles calculation of H@$\beta$-C$_3$N$_2$, F@$\beta$-C$_3$N$_2$, respectively. (g) The Fermi surfaces of H@$\beta$-C$_3$N$_2$ based on single-band Hamiltonian with $E_0$=0 eV, t=-1.52 eV. The charge density distributions in real space of the band crossing the Fermi level are shown in (h) and (i) for H@$\beta$-C$_3$N$_2$ and F@$\beta$-C$_3$N$_2$, respectively.

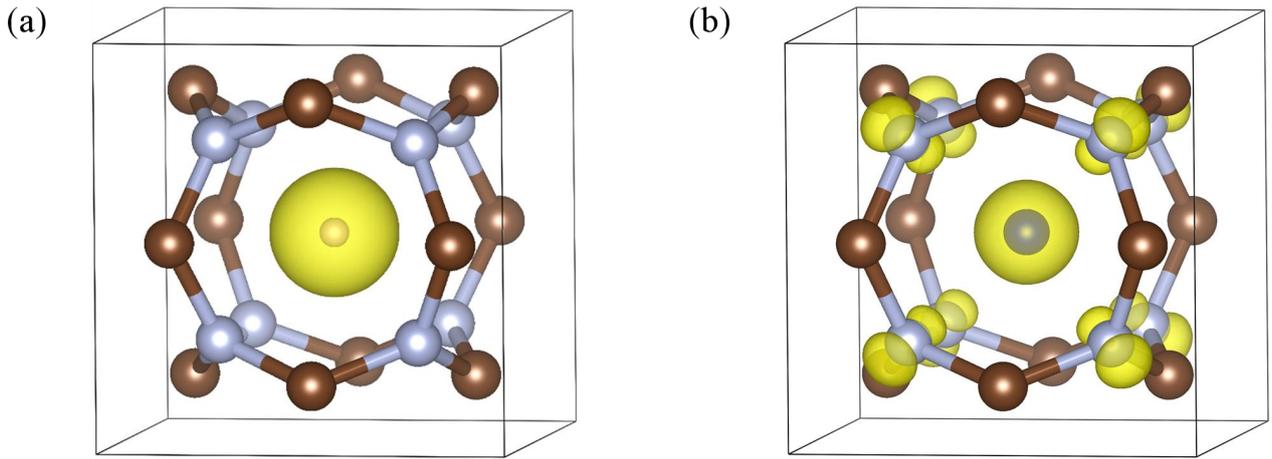

Fig. 3 The spin density of (a) H@$\beta$-C$_3$N$_2$ and (b) F@$\beta$-C$_3$N$_2$

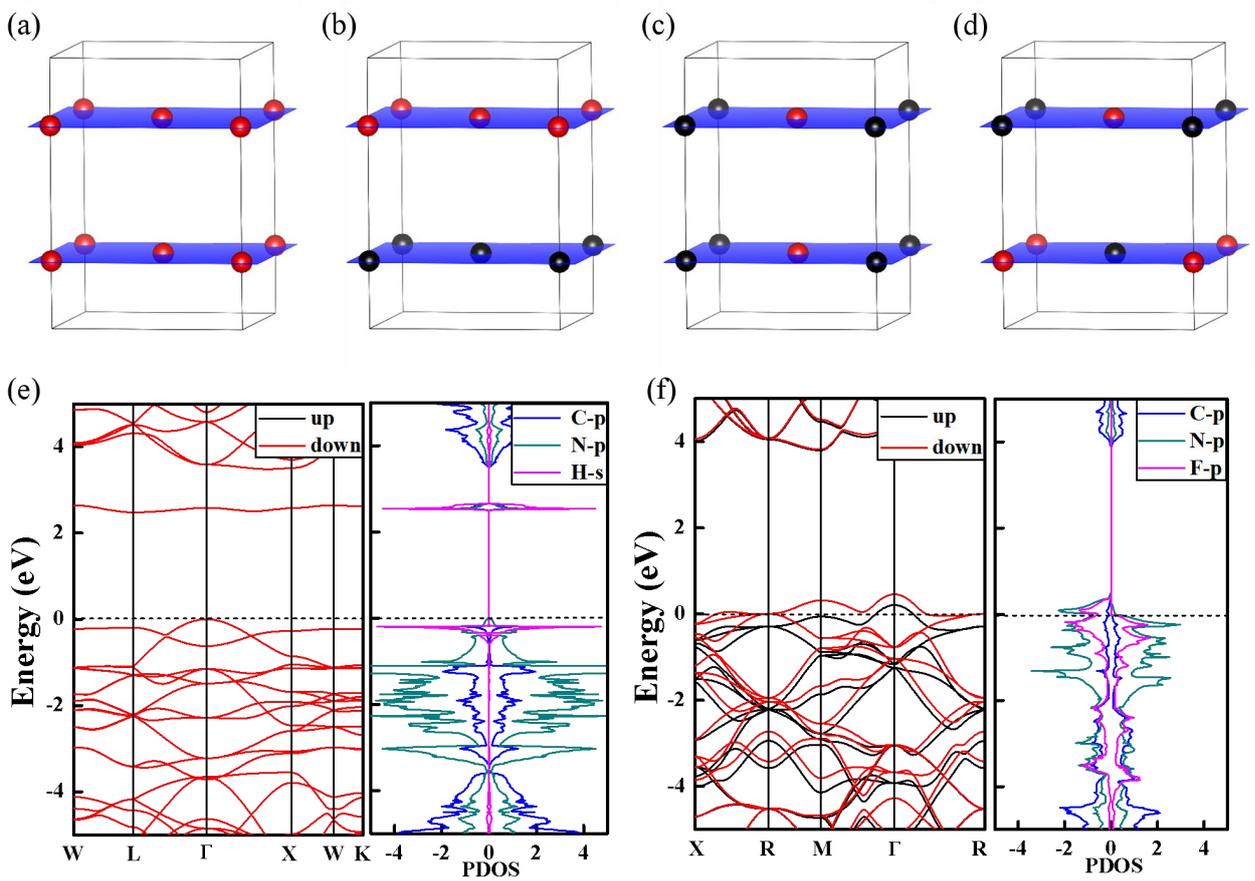

Fig. 4 Four types of magnetic configurations: (a) FM, (b) type-A AFM, (c) type-C AFM, (d) G-AFM. The red and black atoms represent spin up and down, respectively. Band structures and density of states for H@$\beta$-C$_3$N$_2$ with type-G AFM state in (e), and for F@$\beta$-C$_3$N$_2$ with FM state in (f).